\documentclass[twocolumn,amsmath,amssymb,pra,longbibliography,floatfix]{revtex4-1}
\usepackage{graphicx}
\usepackage{hyperref}
\usepackage[none]{hyphenat}
\usepackage{textcomp}

\begin{document}

\sloppy

\title{A universal matter-wave interferometer with optical ionization gratings in the time domain}
\author{Philipp Haslinger}
\author{Nadine D{\"o}rre}
\author{Philipp Geyer}
\author{Jonas Rodewald}
\author{Stefan Nimmrichter}
\author{Markus Arndt}
\email{markus.arndt@univie.ac.at}
\affiliation{University of Vienna, Faculty of Physics, VCQ, Boltzmanngasse 5, A-1090 Vienna, Austria}

\begin{abstract}
Matter-wave interferometry with atoms \cite{Cronin2009} and molecules \cite{Hornberger2012} has attracted a rapidly growing interest over the past two decades, both in demonstrations of fundamental quantum phenomena and in quantum-enhanced precision measurements. Such experiments exploit the non-classical superposition of two or more position and momentum states which are coherently split and rejoined to interfere \cite{Estermann1930,Keith1988,Kasevich1991,Borde1989,Rasel1995a,Fray2004,Mueller2008b, Moskowitz1983,Giltner1995}.
Here, we present the experimental realization of a universal near-field interferometer built from three short-pulse single-photon ionization gratings \cite{Reiger2006,Nimmrichter2011}.  We observe quantum interference of fast molecular clusters, with a composite de Broglie wavelength as small as 275 fm. Optical ionization gratings are largely independent of the specific internal level structure and are therefore universally applicable to different kinds of nanoparticles, ranging from atoms to clusters, molecules and nanospheres. The interferometer is sensitive to fringe shifts as small as a few nanometers and yet robust against velocity-dependent phase shifts, since the gratings exist only for nanoseconds and form an interferometer in the time domain.

\end{abstract}
\maketitle

Recent progress in atom interferometry has been driven by the development of wide-angle beam splitters \cite{Chiow2011}, large interferometer areas \cite{Lan2012} and long coherence times \cite{Mueller2008}. Most interferometers operate in a Mach-Zehnder \cite{Kasevich1991,Keith1991}, $\text{Ramsey-Bord\'{e}}$ \cite{Borde1994} or Talbot-Lau \cite{Clauser1994} configuration, some of them also in the time-domain \cite{Szriftgiser1996,Cahn1997}.  Here we ask how to generalize these achievements to atoms, molecules, clusters or nanoparticles - independent of their internal states.

Mechanical nanomasks \cite{Juffmann2012} could be considered as universal if it were not for their van der Waals attraction on the traversing matter waves, which induces sizable dispersive, that is, velocity-dependent, phase shifts even for gratings as thin as 10 nm.

Optical \cite{Mueller2008b,Chiow2011} or measurement-induced \cite{Storey1992} gratings eliminate this effect, but most methods so far relied on closed transitions and required an individual light source for every specific kind of atom or molecule.

It is possible to circumvent this restriction by using the spatially periodic electric dipole potential in an off-resonant standing light wave. Its field then modulates the phase of the matter wave rather than the amplitude. This implies, however, that the spatial coherence of the incident matter wave needs to be prepared by other means before - such as by collimation, cooling \cite{Deng1999} or the addition of another absorptive (material) mask \cite{Hornberger2012}.

Here, we demonstrate a new method for coherence experiments with a wide class of  massive particles and show how a sequence of ionizing laser grating pulses \cite{Reiger2006}  can form  a generic matter-wave interferometer in the time-domain \cite{Nimmrichter2011}.

\begin{figure*}
\includegraphics[width=0.65\textwidth]{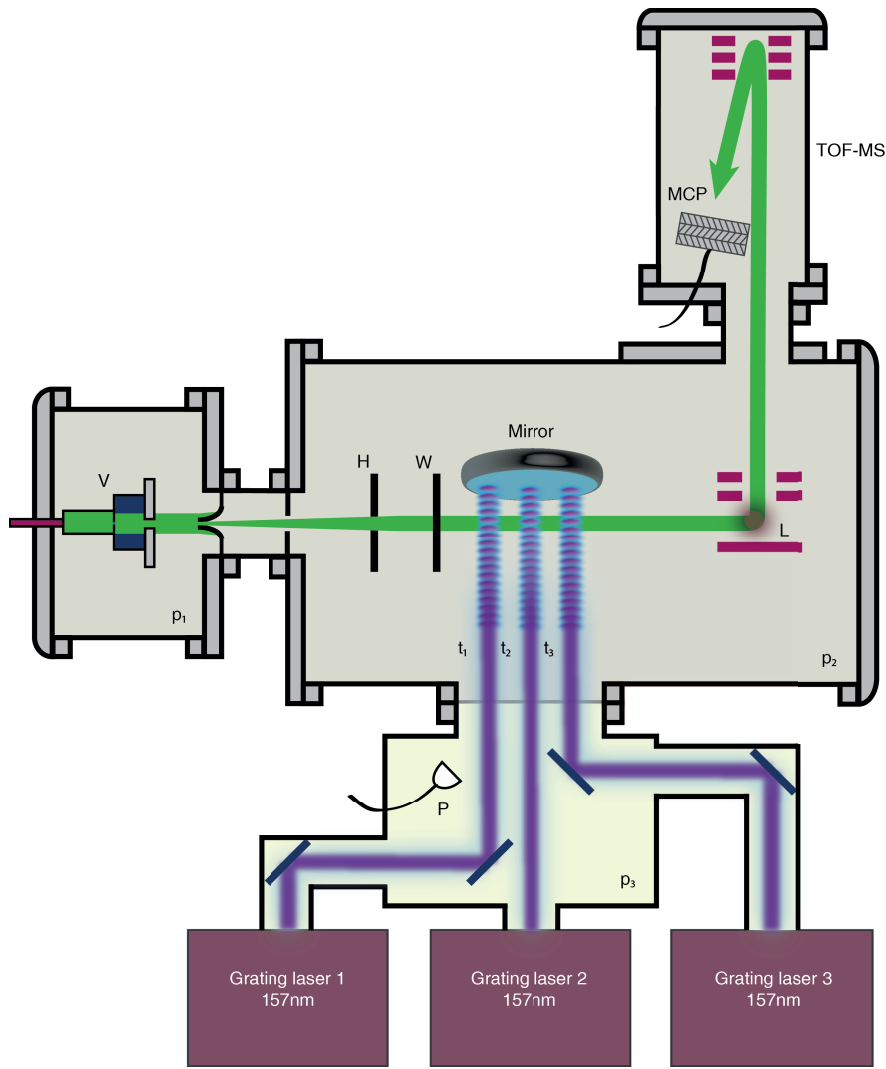}
\caption{\label{fig1}
\textbf{Layout of the OTIMA interferometer}.
\textbf{(a)} Setup for nanoparticle interferometry with three short-pulse optical ionization gratings. From left to right: The Even-Lavie valve (V) produces a 30 \textmu s pulse of neutral anthracene clusters which are cooled in an adiabatic co-expansion with a noble gas jet. The cluster beam is delimited by two slits which are variable in height (H) and width (W). The laser pulses at $\text{t}_1$ = 0, $\text{t}_2$ = T and $\text{t}_3$ = 2T are back-reflected by a single two-inch mirror to form three standing light waves. These are responsible for preparing the initial spatial coherence, for matter-wave diffraction and for spatially filtering the emerging cluster interferogram. The detection laser (L) ionizes the transmitted neutral clusters for time-of-flight mass spectrometry (TOF-MS). A photodiode (P) is used to monitor the laser timing with nanosecond accuracy.
\textbf{(b)} The interferogram is formed by multiple paths from the first to the third grating which correspond to an effective momentum transfer of $\mathrm{n\hbar k}$ in each grating, with n $\in \mathbb{Z}$. Accurate timing ensures that the interfering paths branch and close at the same points on the grating axis x, irrespective of the cluster's initial velocities $\text{v}_1$ (red) $>$ $\text{v}_2$ (green). The stars indicate the localization of the matter waves.
}
\end{figure*}

Figure~\ref{fig1} shows a schematic of the layout of our experiment, which we here realize specifically for clusters of anthracene (Ac) molecules. The molecules are evaporated in an Even-Lavie valve \cite{Even2000} that injects the organic vapor with a pulse width of about 30 \textmu s into the vacuum chamber. The adiabatic co-expansion with a noble gas cools the molecules and fosters the formation of organic clusters - here typically up to $\text{Ac}_{15}$.

The bunch of neutral nanoparticles passes a differential pumping stage, enters the interferometer chamber and flies in a short distance (0.1 - 4 mm) from the surface of a super-polished $\text{CaF}_{2}$ mirror before it reaches the laser ionization region of a time-of-flight mass spectrometer (ToF-MS) where it creates the signal peaks.

The pulsed beams of three synchronized $\text{F}_2$-excimer lasers ($\lambda$ = 157.63 nm) hit the mirror surface and the cluster beam under normal incidence with a variable pulse energy of 1 - 3 mJ and a duration of about 7 ns.  The laser beams are separated in space by $\sim$ 20 mm along the cluster trajectory. Their mutual time delay is adjusted with an accuracy of a few nanoseconds. We choose the laser beam diameters ($\sim$ 1 mm $\times$ 10 mm rectangular flat top, extended along the cluster beam) to cover a wide particle bunch emitted by the source, whereas the detection laser beam is narrow enough to post-select only those clusters that have interacted with all three laser light pulses.

\begin{figure*}
\includegraphics[width=\textwidth]{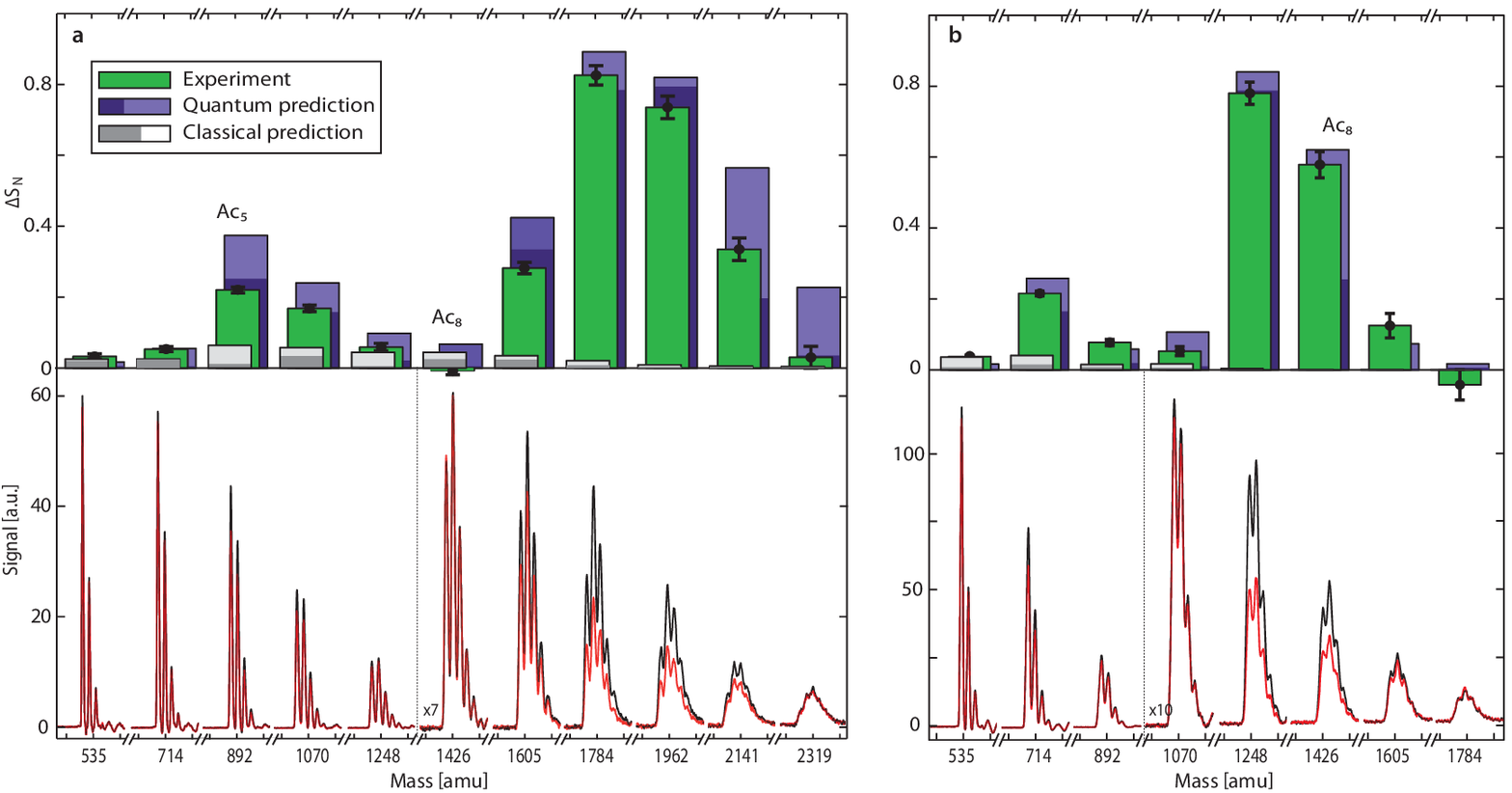}
\caption{\label{fig2}
\textbf{Cluster interference visualized via the mass spectrum, for two pulse separation times.}
\textbf{(a)} Bottom panel: Mass spectra recorded for a resonant (black line) and off-resonant ($\Delta$T = 200 ns, red line) pulse separation of T = 25.9 \textmu s (clusters seeded in an argon jet).  Each cluster signal splits in isotopic sub-peaks. The x-ticks correspond to a mass separation of 4 amu. The two spectra differ for masses which fulfill $\text{T}_\text{m} \simeq$ T. Top panel: Histogram of the cluster interference contrast, as measured by the signal difference $\Delta \text{S}_\text{N}$ integrated over the main isotopes of a given cluster. The predictions of the quantum/classical model \cite{Nimmrichter2011} are shown in violet/grey. The light violet/grey regions indicate the variation of the fringe contrast with a $\pm 30\%$ variation of the cluster polarizability $\alpha_{157}$. For further details, see Methods and Suppl. Inf. e \textbf{(b)} Same as \textbf{(a)} but with neon seeding and T = 18.9 \textmu s. The error bars represent one standard deviation of statistical error (see Suppl. Inf. h).
}
\end{figure*}

All three laser gratings interact with the matter waves in two different ways \cite{Nimmrichter2011}: they imprint a periodic phase and, more importantly, they act as transmission gratings because the photon energy of $\sim$ 7.9 eV exceeds the ionization energy of the nanoclusters. Particles that traverse the antinodes of a laser grating ionize with high probability after absorption of one or more photons and a weak electric field removes them from the beam. Close to the nodes of the standing light waves the clusters remain neutral and move on in the interferometer. This process imprints a periodic modulation onto the matter-wave amplitude - as if the clusters had passed a mechanical nanomask.

A strong spatial localization inside the first laser grating is important for preparing a comb of emergent wavelets whose transverse coherence will cover a few antinodes in the second light grating further downstream. This is a prerequisite for interference to occur, that is, for the formation of a free-flying cluster density pattern at precisely defined moments in time, which is probed with nanosecond precision by the third ionizing standing wave.

The three laser pulses form a Talbot-Lau interferometer in the time domain, which exhibits transmission resonances when the delay between two subsequent pulses is close to the Talbot time $\text{T}_\text{m} = \text{md}^{2}\text{/h}$, with m the cluster mass and h Planck's constant. In our setting the grating period d =$\lambda$/2 = 78.8 nm results in $\text{T}_\text{m}$ = 15 ns/amu.  All particles see the same gratings at the same time irrespective of their velocity.  Even though they may enclose different areas in real-space ($x-z$), they will accumulate the same phase and contribute constructively to the same interferogram for each given mass (Figure 1b).

We trace the emergent interference pattern in four different ways: its mass characteristics, its dependence on the pulse separation and pulse sequence asymmetry, and by visualizing its structure in position space.

\begin{figure}
        \includegraphics[width=\columnwidth]{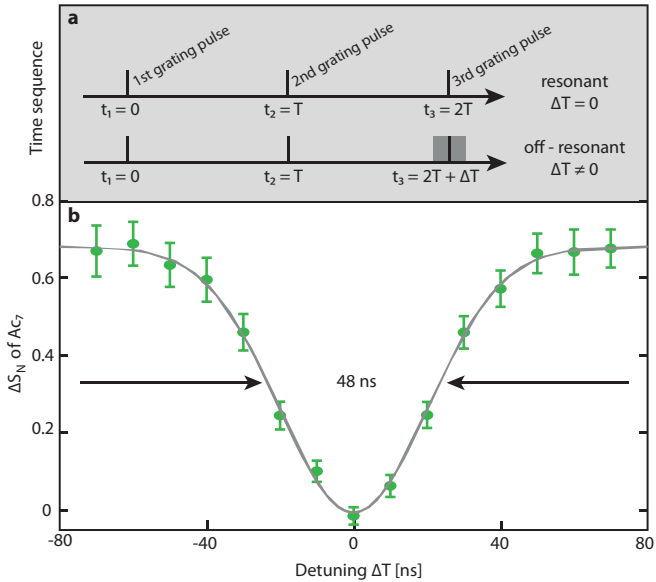}
        \caption{\label{fig3}
        \textbf{Interferometric resonance and timing precision.}
        Cluster self-imaging in a pulsed near-field interferometer is a resonant process with a short acceptance window for the matter waves to rephase. \textbf{(a)} Pulse sequence and \textbf{(b)} difference $\Delta \text{S}_\text{N}$ between the resonant and off-resonant signals detected at a mass of $\text{Ac}_7$ as a function of $\Delta$T. In our setup and for a pulse separation time T of 18.9 \textmu s interference occurs during a time window of 48 ns (FWHM). The error bars represent one standard deviation of statistical error (see Suppl. Inf. h).
        }
\end{figure}

We start by monitoring the ToF-MS signal and toggle between a resonant and a non-resonant setting.  In the resonant mode the delays $\text{t}_2-\text{t}_1$ = T,  $\text{t}_3-\text{t}_2$  = T + $\Delta \text{T}$ between two subsequent laser pulses are equal, $\Delta$T = 0, and quantum interference is expected to modulate (enhance or reduce, depending on the phase) the transmission for the mass whose Talbot time matches the pulse separation T.  In the off-resonant mode, the pulse delays are imbalanced by $\Delta$T = 200 ns and this tiny mismatch suffices to destroy the interferometric signal. We extract the interference contrast from the normalized difference $\Delta \text{S}_\text{N} = (\text{S}_\text{R} - \text{S}_\text{O}) / \text{S}_\text{O}$ between the resonant $\text{S}_\text{R}$ and the off-resonant signal $\text{S}_\text{O}$ and plot it as a function of mass in Figure~\ref{fig2}. The experimental mass spectra  and $\Delta \text{S}_\text{N}$ bars (green) can be well understood by a quantum mechanical model (violet bars), as described in the Methods Section, and both are in marked discrepancy with a classical model (grey bars) \cite{Nimmrichter2011}.

The role of the pulse separation T is demonstrated by changing the seed gas from argon to neon. Shifting the most probable jet velocity from 690 to 925 m$\text{s}^{-1}$ allows us to decrease T. The quantum model then predicts the highest contrast to occur at smaller masses, as confirmed by the experimental data in Figure 2b.

Figure~\ref{fig3} shows a clear resonance in $\Delta \text{S}_\text{N}$ as a function of the time imbalance $\Delta \text{T} \in$ [-70, +70] ns with a width determined by the transverse momentum distribution of the cluster beam \cite{Nimmrichter2011}. The momentum spread inferred from a Gaussian fit to the data in Figure~\ref{fig3} corresponds to a divergence angle along the grating of 2.1 mrad, in good agreement with the experimental settings.

In our set-up, the pulsed supersonic expansion determines the cluster velocity distribution and the pulsed mass detection post-selects its relative width to $\Delta$v/v $\simeq$ 3 $\%$. It is then justified to interpret the observations in position space: With the de Broglie wavelength given by
$\lambda_{\text{dB}}$ = h/mv the mass distribution also represents a wavelength spectrum. The most prominent interference peak in Figure 2b at 1248 amu corresponds to the heptamer $\text{Ac}_7$ with $\lambda_{\text{dB}}$ $\simeq$ 345 fm, at v $\simeq$ 925 m$\text{s}^{-1}$. The highest mass peaks in the spectrum reach down to below $\lambda_{\text{dB}}\simeq$ 275 fm.

\begin{figure}
        \includegraphics[width=\columnwidth]{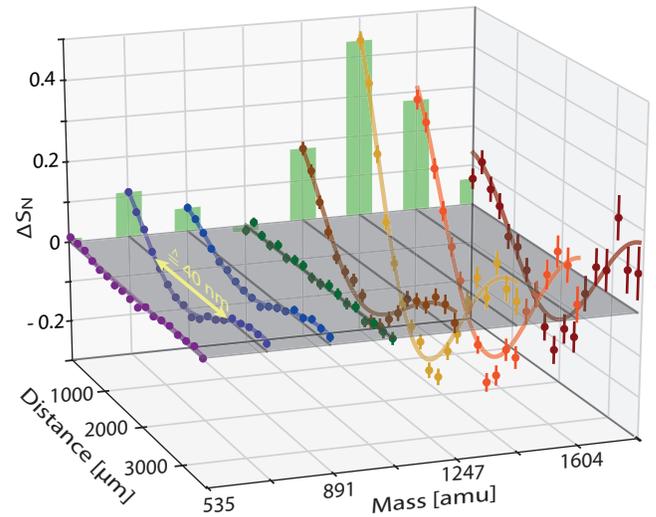}
        \caption{\label{fig4}
        \textbf{$\Delta \text{S}_\text{N}$ as a function of the mirror displacement for different clusters.}
        The second grating laser beam was tilted by 5.1 $\pm$ 0.3 mrad in the direction of the molecular beam to stretch the effective grating period by about 0.013 per mille. This suffices to induce a fringe shift of half a grating period for molecules travelling around 1.5 mm distance from the mirror surface. The mirror height is varied to effectively shift the second grating with regard to the other two which allows us to scan the cluster interference pattern.  We extract the periodicity for $\Delta \text{S}_\text{N}$ as a function of the mirror distance by fitting a damped sine curve to the experimental data.  This periodicity corresponds to the expected effective period \cite{Nimmrichter2011} of the interferogram of 80 nm. The error bars represent one standard deviation of statistical error (see Suppl. Inf. h).
        }
\end{figure}

Finally, we can also prove the formation of an interference pattern in real space by modifying the period of the central grating: While all laser beams had originally been set to normal incidence on the interferometer mirror - with an uncertainty of about 200 \textmu rad - we now explicitly tilt the central laser beam by 5.1 mrad along the cluster beam. The direction of the standing-light-wave grating remains defined by the orientation of the mirror surface, but an increasing tilt angle $\theta$ reduces the modulus of the wave vector perpendicular to the surface, $\text{k}_\text{p} \simeq \text{k}\cdot\cos \theta$. We can shift the interference pattern by half a grating period when the clusters pass the mirror surface at an average distance of 1.5 mm. We plot the fringe shift as a function of the separation between the beam and the mirror in Figure~\ref{fig4} and find a damped sinusoidal transmission curve for all clusters with the expected period. The overall damping results from the limited coherence of the laser system and the vertical extension of the Ac cluster beam.

All tests presented here confirm the successful experimental realization of an \textbf{o}ptical \textbf{t}ime-domain \textbf{i}onizing \textbf{ma}tter-wave (OTIMA) interferometer \cite{Hornberger2012,Nimmrichter2011}, which exploits pulsed ionization gratings. This versatile tool for quantum interferometry will be applicable to a large class of nanoparticles.
Owing to the pulsed gratings, all phase shifts caused by constant external forces become velocity-independent and leave the contrast unaffected. The dispersive Coriolis shift \cite{Lan2012} can be well compensated by a suitable orientation of the interferometer, if needed.

The wide applicability and non-dispersive nature of pulsed ionization gratings make the OTIMA design particularly appealing for quantum experiments with highly complex particles, eventually even with nanoparticles at the length scale of the grating period.  As high-mass interferometry requires coherence of the order of the Talbot time, practical mass limits are imposed by free fall in the gravitational field on Earth in combination with the limited coherence of vacuum ultraviolet lasers and the finite phase-space density of the available particle sources. However, none of them is fundamental. Even in the presence of thermal radiation at room temperature (particle and environment) and collisional decoherence at a background pressure of $10^{-9}$ mbar, the OTIMA design is predicted to enable new tests of quantum physics, such as tests of spontaneous localization, with particle masses around $10^6$ amu and beyond \cite{Nimmrichter2011b}.

On the applied side, the OTIMA set-up is expected to improve the accuracy of  molecule and  cluster deflectometry because it ensures the same interaction (phase accumulation) time for all particles with the external fields \cite{Heer2011} and a position readout at the nanometre scale. Our interferometer concept therefore establishes also the basis for a new class of quantum-enhanced precision metrology experiments.

\section*{Methods}

\emph{Absorption and optical polarizability.}
The central grating influences the propagation of the coherent matter wave by modulating both its amplitude and phase. It does this by removing particles from the anti-nodes of the standing light field and by imprinting a phase onto the matter wave in proportion to the clusters' optical polarizability at 157 nm.  In the first and third grating the phase modulation has no effect, since the clusters enter with random phases, and since the last grating merely acts as a transmission mask.  Neither the absorption cross sections $\sigma_{157}$(N)  nor the polarizabilities $\alpha_{157}$(N) are known, a priori, for each cluster of N molecules in the vacuum ultraviolet wavelength range. However, $\sigma_{157}$(N) enters the model only through the mean number of photons absorbed $\text{n}_0\left(\text{N}\right)$ in each grating which we can determine by monitoring the cluster loss rate. While this parameter influences the general shape of the interference curve as a function of mass, the polarizability may modify the predicted contrast of each individual cluster.  We assume the polarizability and the absorption cross section to exhibit the same N-scaling as retrieved from our $\text{n}_0\left(\text{N}\right)$ measurements and we allow the polarizability to vary by $\pm$ 30 $\%$ (light violet confidence areas in Figure~\ref{fig2}) around the single-molecule value. We use the polarizability  $\alpha_{157}$(1) = 25.4 $\times 10^{-30}\,\text{m}^3$ from Marchese et al. \cite{Marchese1977} and we extract an absorption cross section of $\sigma_{157}$(1) = 1.1 $\times 10^{-20}\,\text{m}^2$ from Malloci et al. \cite{Malloci2004}.

This yields the quantum and classical theory curve in Figure~\ref{fig2}. Apart from the uncertain polarizability, the deviations from the experimental data may be attributed to a limited efficiency of single-photon ionization and contributions by fragmentation processes.  While the absolute interference contrast is sensitive to a variety of different cluster properties which still wait to be extracted in combination with more refined cluster theory, the fringe shift will become valuable for precisely measuring the interplay between internal cluster properties and external forces.

\section*{acknowledgments}
We acknowledge support by the Austrian science funds (FWF-Z149-N16 Wittgenstein and DK CoQuS W1210-2) as well as infrastructure funds by the Austrian ministry of science and research BMWF (IS725001). We thank Uzi Even and Ori Cheshnovsky for emphasizing the benefits of organic molecules with the Even-Lavie valve and Klaus Hornberger for collaborations on the modeling of the OTIMA interferometer.  We thank Bernd von Issendorff for discussions on cluster sources.

\bibliography{otimaref}

\section*{Supplementary Information}

\subsection*{(a) Source}
We use an Even-Lavie (EL) valve to create a pulsed neutral molecular cluster beam. Anthracene (Ac) molecules are heated close to their melting point (491 K) in the valve and they are co-expanded into high vacuum with a supersonic noble gas jet (p $\simeq$ 1 - 10 bar). There they cool and condense to form clusters ranging from $\text{Ac}_1$ - $\text{Ac}_{15}$. The EL valve is operated at a repetition rate of 100 Hz and it is synchronized with the three vacuum ultraviolet (VUV) grating lasers and the detection laser.

\subsection*{(b) VUV Laser system}
The gratings are generated by three synchronized GAM lasers, model EX50F, $\simeq$ 5 mJ, shot-to-shot energy stability $\simeq$ 5$\%$, coherence length $\simeq$ 1 cm. The grating transmission function depends on the laser energy which we monitor for every individual pulse via the photodiode P (in Figure~\ref{fig1}). All mass spectra are sorted according to a given laser energy and pulse delay $\Delta$T. The laser beam lines are evacuated and purged with dry nitrogen (6.0) at 1 mbar to avoid both absorption and laser-induced deposition of debris on the mirrors. The ionizing VUV laser at the TOF-MS is a Coherent Excistar $\text{F}_2$-laser (3 mJ) with an energy stability of better than 5$\%$.

The $\text{F}_2$ laser operates mainly at 157.63 nm, with an additional weak line at 157.52 nm \cite{Sansonetti2001}.  Our own measurements on the GAM lasers confirm the specified coherence length of $\text{l}_c \simeq$ 1 cm which ensures a standing wave in a few millimeters distance to the mirror. The transverse coherence of each excimer laser is given by its output aperture and amounts to $\simeq$ 40 $\mu$m at the mirror.

The timing sensitivity of the OTIMA scheme at short pulse separation periods requires a precise monitoring of the laser jitter. The intrinsic short term jitter of all three GAM lasers is less than 7 ns (FWHM). They exhibit, however, long term instability of the order of 100 ns, which we measure and compensate. Our Coherent laser jitters by $\simeq$ 20 ns (FWHM). The timing of the grating laser pulses is recorded to post-select the interferograms according to their pulse-delays. All measurements were made with a maximal jitter smaller than 5 ns.

\subsection*{(c) VUV mirror}
The dielectric interferometer mirror (Jenoptik, Germany) is made from VUV grade $\text{CaF}_2$ coated with a reflectivity of R $>$ 96$\%$ under normal incidence.  The finite reflectivity allows us to monitor the position and shape of the laser beams via their scattering on the frosted backside of the mirror.

Moreover it causes a small running wave to add to the standing wave. The constant intensity offset would only slightly reduce the interference contrast if every cluster was always ionized by a single photon. Our own and independent measurements indicate a minimal spherical deformation across the two-inch mirror towards its edges up to 100 nm.

\subsection*{(d) Mass spectrometry}
The time of flight mass spectrometer (ToF-MS, Kaesdorf Munich) is built as an orthogonal reflection MS with $\Delta$m/m = 1/3000. The relative mass spread across every individual multiplet is as small as 0.1 - 1 $\%$. A mass variation of $\pm$ 2 amu on m = 1400 amu gives rise to a variation of 30 ns on 20 \textup{$\mu$}s Talbot time. This leads only to a negligible reduction of the interference contrast.

\subsection*{(e) Number of absorbed photons}
We chose anthracene as a test molecule because its ionization energy $\text{E}_\text{i}$ is smaller than the photon energy at 157.63 nm (7.9 eV) and the contrast is highest if the absorption of a single photon suffices to ionize the particle. If this condition is fulfilled for a certain cluster number N it will be generally met for all higher clusters too, since $\text{E}_\text{i}$ decreases with cluster size to approach the work function of the bulk. Photon absorption without subsequent ionization would diminish the interference contrast. The photoionization quantum yield \cite{Jochims1996} of anthracene at 7.9 eV is only ~10 $\%$. Our data are compatible with the assumption that it is close to one for clusters composed of several molecules.
Different structural isomers may respond differently to the incident light, but a full assessment of all optical properties for all cluster sizes is beyond the scope of this first demonstration of experimental OTIMA interferometry.

\subsection*{(f) Vacuum system}
The source chamber is evacuated to $\text{p}_1$ = 1 $\times 10^{-5}$ mbar, the interference chamber to
$\text{p}_2 = 2 \times 10^{-8}$ mbar and the optical beam line to $\text{p}_3$ = 1 mbar.

\subsection*{(g) Data Recording and Processing}
The TOF-MS voltage signal is recorded using a 10 bit digitizer card (Agilent Acquiris DC282) with 0.5 ns time resolution.  We run the experiment with 100 Hz repetition rate. A data file for one mass spectrum has a size of 1 mega points. Data are post-processed in real time using a custom developed software solution. The software also records the laser timings and pulse energies.

\subsection*{(h) Figures}
\textbf{Figure~\ref{fig2}:} The TOF-MS data were averaged over about 28000 individual mass spectra for panel (a) and about 14500 spectra for panel (b). An overall TOF-MS background was subtracted, for all masses equally.
The green columns in the upper panels of Figure 2a and 2b were generated by summing the mass spectra (bottom panels) over a mass region whose width is indicated by the width of the columns. It accounts for the majority of the isotopic spread of a given cluster peak.
The experimental error bar was determined as follows: Since the experimental response to the incidence of an ion is a voltage peak whose amplitude changes both from shot to shot and with increasing ion mass, we chose to extract a measure for the true count rate from the observation of ``no count'' - a small discriminator threshold was set to distinguish between the presence or absence of ions - in every mass bin. Assuming a Poissonian distribution of the cluster counts one can then infer the average detected cluster number and its standard deviation ($\delta \text{S}_\text{O}$ and  $\delta\text{S}_\text{R})$   from  the probability of finding zero counts. The error of the normalized signal difference $\delta\left(\Delta \text{S}_\text{N}\right)$ is then computed using Gaussian error propagation.
The data has been evaluated and plotted with Matlab R2010b and arranged using Adobe Illustrator CS5.
\\

\textbf{Figure~\ref{fig3}:} For this data set, TOF-MS data were averaged over 3300 - 3500 frames for each data point. The error bar was determined by the same procedure as in Figure 2. The data has been evaluated and plotted with Matlab R2010b and arranged using Adobe Illustrator CS5.
\\

\textbf{Figure~\ref{fig4}:} For this data set we averaged over roughly 25000 mass spectra for every step in mirror distance.  Uncertainty bars were generated using the same procedure as in Figure 2 and 3. The data has been evaluated with Matlab R2010b, plotted using the Matplotlib package for Python and arranged using Adobe Illustrator CS5.

\end{document}